\documentclass{article}                  
\usepackage{graphicx}
\usepackage{subfigure}
\usepackage{amssymb}
\usepackage{multicol}
\usepackage{amsmath}
\usepackage{lipsum}  
\usepackage{cite}
\usepackage{textcomp}
\def\BibTeX{{\rm B\kern-.05em{\sc i\kern-.025em b}\kern-.08em
    T\kern-.1667em\lower.7ex\hbox{E}\kern-.125emX}}
    
 \usepackage{datetime}
    
\usepackage{color}

\providecommand{\com[2]}{\begin{tt}[#1: #2]\end{tt}}

\newcommand{\beq}{\begin{equation}}
\newcommand{\eeq}{\end{equation}}
\newcommand{\beqa}{\begin{eqnarray}}
\newcommand{\eeqa}{\end{eqnarray}}
\newcommand{\beqan}{\begin{eqnarray*}}
\newcommand{\eeqan}{\end{eqnarray*}}
\newcommand{\bef}{\begin{figure}}
\newcommand{\enf}{\end{figure}}

\newcommand{\bi}{\begin{itemize}}
\newcommand{\ei}{\end{itemize}}
\newcommand{\bc}{\begin{center}}
\newcommand{\ec}{\end{center}}
\newcommand{\ba}{\begin{array}}
\newcommand{\ea}{\end{array}}
\newcommand{\be}{\begin{equation}}
\newcommand{\ee}{\end{equation}}
\newcommand{\beno}{\begin{equation*}}
\newcommand{\eeno}{\end{equation*}}
\newcommand{\beqna}{\begin{eqnarray}}
\newcommand{\eeqna}{\end{eqnarray}}
\newcommand{\bd}{\begin{displaymath}}
\newcommand{\ed}{\end{displaymath}}
\newcommand{\beqnd}{\begin{eqnarray*}}
\newcommand{\eeqnd}{\end{eqnarray*}}

\newtheorem{problem}{Problem}

\definecolor{red}{rgb}{1,0,0}

\definecolor{blu}{rgb}{0,0,1}

\definecolor{gre}{rgb}{0,0.7,0.3}

\begin{document}

\title{Improving the GMAW process through current control} 
 
\author{Alexandre Sanfelici Bazanella\thanks{A. S. Bazanella and M.G. de Freitas are
with the Data-Driven Control Group, Department of Automation and Energy, Universidade Federal do Rio Grande do Sul,
Porto Alegre-RS, Brazil. Email: bazanella@ufrgs.br. This study was financed by Conselho Nacional de Desenvolvimento Cient\'{\i}fico e Tecnol\'ogico (CNPq)} \and
Mateus Gaspary de Freitas\thanks{M.G. de Freitas is with SUMIG - Solutions for Cutting and Welding, Caxias do Sul - RS, Brazil. Email: mateus.gdfreitas@gmail.com}}

\date{\empty}

 \maketitle

 \begin{abstract}
A control strategy for the electrical current in GMAW processes is proposed.  The control is in closed-loop, designed by formal methods, based on a mathematical model of the electrical behavior of the GMAW process, and implemented in C+ language in a microcontroller. The model consists of a switched equivalent electrical circuit whose parameters are obtained in a data-driven manner. The strategy is tested in numerous experiments with both manual and robot welding, showing improvements in the overall welding process. 
  \end{abstract}

 \section{Introduction}
 
%


In the Gas Metal Arc Welding (GMAW) process, there is a strong relationship between the quality of the process and the waveform of the electric current in each phase of the process \cite{Naidu:2003}. Thus a precise control of current, so that a particular (optimal) waveform is enforced with higher precision, has strong potential to improve the overall quality of the process and its results. Accordingly, there is continuous effort both academic and industrial, to better understand this relationship between the current waveform and welding quality, as well as to improve the control of current. Recent developments along this line include \cite{Silva:2024, Chaturvedi:2021, corrente1,corrente2,corrente3, PID}; a review is given in \cite{Dutra:2023}. 

In this paper we propose a particular control strategy for the electrical current. A closed-loop switched PID (Proportional-Integral-Derivative) controller is formally specified, designed (tuned) by formal methods and implemented in a microprocessor. The control specifications are based on the desired current waveform and the control design is based on an equivalent circuit model. This model is conceptual, but we propose to perform a data-driven identification of its parameters, so that it becomes also  quantitatively meaningful and thus useful for simulation and for control design. The switched PID controller is then tuned with the Root Locus method to satisfy the specifications. 

Numerous experimental results with the application of the proposed controller are presented, involving both manual and robot welding. The results illustrate various aspects of the process: experimental current and voltage waveforms are shown along with quantitative performance measures of their quality; the welders' qualitative assessment is reported; metallographic analyses of various samples are presented and discussed. All the results support the overall high quality of the process obtained with our proposed control strategy. 

 \section{The GMAW process}
 
 \subsection{General principles}
 
\begin{figure}[ht!]
    \centering
         \caption{Schematics of the GMAW process}
        \includegraphics[scale=0.35]{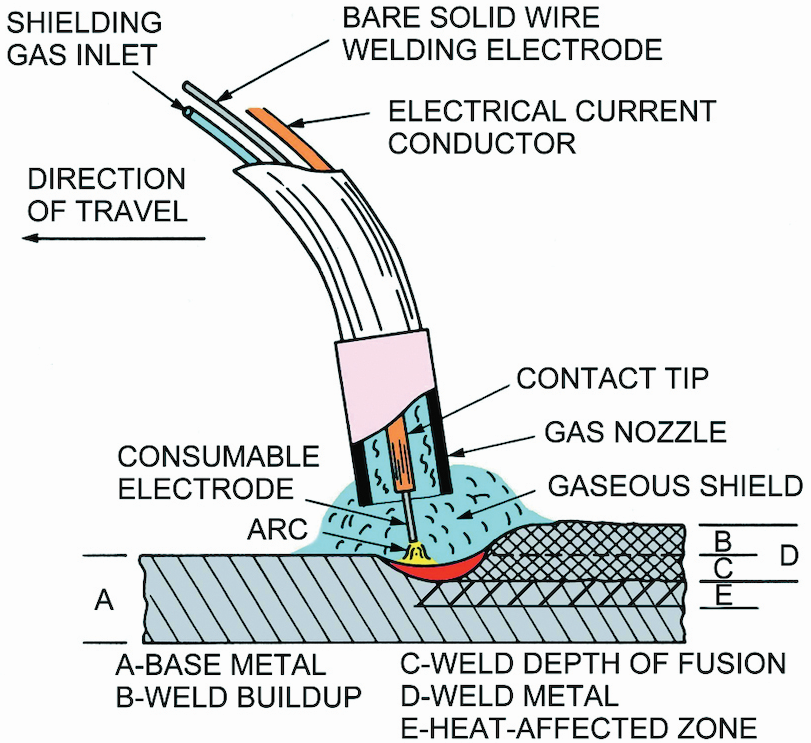}\\
        \label{fig:1}
\end{figure}

The GMAW process uses an electric arc as a heat source to join two metal pieces. The process occurs between a continuously fed consumable electrode wire and the molten pool, protected by an external gas, without the application of pressure \cite{AWS:2001}. The consumable electrode wire is fed continuously through a conduit guide, with the aid of a traction or driving system, usually driven by a motor. This wire is directed to the nozzle of a welding torch, where it is energized by the contact tip, effectively becoming an electrode.  An electric arc is formed when the energized electrode comes into contact with the workpiece to be welded, which is energized in the opposite way, generating the passage of high current due to a short circuit. The electric arc is maintained by constantly feeding the wire and by the control of the welding source, while the welder - whether a welding professional or robot - is responsible for activating, deactivating and moving the torch around the joints. Figure \ref{fig:1} illustrates the schematic diagram of the GMAW welding process. 

 The electric arc transforms the continuously fed electrode wire into drops of molten metal,
 which are transferred to the workpiece, forming a weld bead. 
Controlling the balance between the wire feed speed and its fusion, in order to guarantee a metal transfer that ensures high quality, productivity and repeatability, is a very complex task. 
The complexity of the physical phenomena involved in metal transfer and the
electric arc require a detailed understanding for process optimization, making its continuous improvement essential for welding efficiency.

\begin{figure}[ht!]
    \centering
         \caption{Metal transfer cycle in the  GMAW process}
        \includegraphics[scale=0.4]{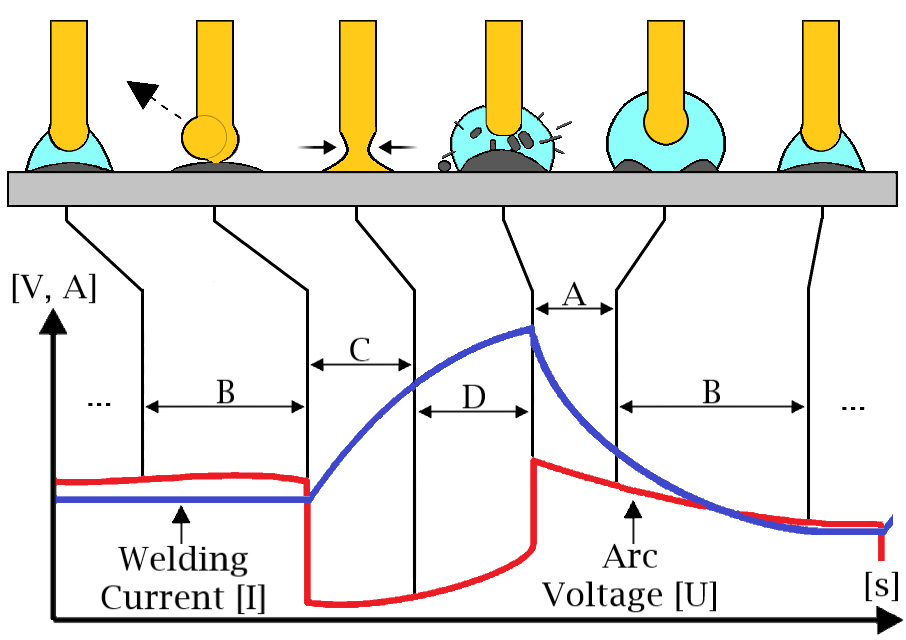}\\
        Adapted from \cite{Silva:2024}.
        \label{fig:metal_transfer}
\end{figure}

 Short-circuit transfer in the welding process occurs when the liquid metal drop, still connected to the electrode wire, comes into direct contact with the molten pool. Initially spherical, the drop elongates under the action of the forces involved, forming a cylindrical metal bridge that stabilizes when it reaches the diameter of the wire. The constriction of this bridge progressively intensifies until it collapses, transferring the drop to the molten pool and reestablishing the arc. This cycle repeats itself continuously, characterizing the dynamics of the process \cite{Silva3:2008}.
 Figure \ref{fig:metal_transfer} shows the stages and typical waveforms of voltage and current during the process. 
In stages C and D the contact is in short-circuit; accordingly, this is called the {\em short-circuit phase}.
It is seen in Figure \ref{fig:metal_transfer} that during the short-circuit phase the current grows almost exponentially,
 as in the charge of a passive electrical circuit.
 In stages A and B the tip of the moving wire is closed to the substrate, but separate from it and a strong electric arc
is formed; accordingly, this is called the {\em electric arc phase}. During the electric arc phase, the voltage between
the wire and the substrate grows as the current passing through the arc is reduced, in a similar way to 
the discharging of an electrical  circuit. Melting of the wire occurs mainly during the electric arc phase,
whereas deposition of the melted metal onto the substrate occurs mainly during the short-circuit phase.
As seen in the archetypal curves in Figure \ref{fig:metal_transfer}, in each phase the electrical characteristics are different.

%
%

 
 \subsection{Electrical characteristics}
 
Current flowing through the arc is what makes welding happen. So, it is not surprising that current behavior determines to a large extent the characteristics of the process and thus its quality. An erratic behavior of current can result in an erratic melting and, as a consequence, in an unpredictable quality of the final joint. On the contrary, particular current waveforms, if properly and precisely enforced, provide  optimal conditions for welding. 

Industrial practice corroborates this natural expectation. It has been found through industrial experience that indeed the current behavior is highly correlated to the process' quality. A particular waveform that provides excellent results is when the current going through the arc grows at an optimal constant rate during the short-circuit phase and decays at another optimal constant rate during the electric arc phase. These optimal rates depend on the materials being fused and on the gas \cite{Dutra:2023, Silva:2024, corrente2}.

A commonly applied solution in industrial practice to obtain such a behavior is to include an inductor in the loop  to ``control'' the current's rate. This solution can be justified by the following rationale, which is convenient to motivate and justify our design approach. The current flows through an electrical circuit with constant voltage $V_0$ whose behavior can be described by a differential equation in the form 
\begin{equation}\label{ED}
L\frac{di(t)}{dt} + o(t) = V_0
\end{equation}
where $L\frac{di(t)}{dt}$ is the voltage drop at the inductance and $o(t)$ collects the voltage drops at all other components in the circuit. If the inductance is large enough, then the first term in the left-hand side of \eqref{ED} dominates the second one, the voltage drop concentrates mostly on the inductance and the derivative of the current can be approximated by 
\begin{equation}\label{L}
\frac{di}{dt} \approx \frac{V_0}{L}.
\end{equation}

One thus picks an inductance such that $\frac{V_0}{L}$ gives the desired rate. This practice can be interpreted as an open-loop control policy, justified by a radically simplified model (equation \eqref{L})  and implemented by picking the characteristic of a physical component of the electric circuit - the inductance. This procedure usually ensures that the derivative of the current stays around the desired value most of the time, but clearly it does not make the current to behave according to the specification, that is, with a constant derivative at each stage. Not only this solution is obviously not optimal and imprecise; it is also very expensive to procure an inductance to endure such a high current level. In fact, nowadays the inductances represent a large portion of the cost of a MIG/MAG machine. 
A more precise control of current has the potential of improving significantly the quality of the welding. If implemented properly, that is, by means of closed-loop current feedback, it will also reduce significantly the cost by eliminating the need for a large inductance.
In order to design such a controller, an electrical  model of the GMAW process will be presented in the next Section.

\subsection{Experimental setting}

\begin{figure}[ht!]
    \centering
         \caption{SUMIG Intellimig500 inverter welding power source}
        \includegraphics[scale=0.43]{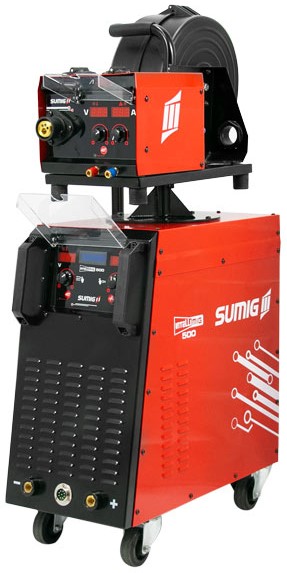}\\
        \label{fig:WeldingPS}
\end{figure}

All the experiments were performed on an industrial welding equipment: the SUMIG inverter welding power source, Intellimig500, illustrated in Figure \ref{fig:WeldingPS}, was used. This equipment has a nominal apparent power of 14.4 kVA, a power factor of 0.86, and is capable of supplying up to 40 V and 500 A for the GMAW welding process. Featuring a constant-voltage characteristic and continuous wire feeding, its feeding speed ranges from 0.8 to 22 m/min, making it suitable for conventional GMAW welding applications. 

The \textit{Intellimig500} welding source uses the STM32F407VET6 microcontroller from STMicroelectronics\texttrademark, which features an Arm$^{\circledR}$ 32-bit Cortex$^{\circledR}$ -M4 CPU architecture, 12-bit analog-to-digital converters, and a processing frequency of up to 168 MHz, allowing control of the entire welding process.
Our proposed current control strategy was programmed in C++ language in
this microcontroller, thus reprogramming the welding source.

 \section{Modeling}
 
 Welding is a highly complex process whose analysis has received a lot of attention in the literature. Such analysis efforts address the modeling of various aspects of the process. In order to control the current through the joint, it is the electrical characteristics of the circuit generating the electric arc that are relevant to model. The voltage applied to this circuit, often kept constant during the welding process, is the natural variable to be manipulated to control the current. In this Section we propose a modeling process for this electrical behavior, which will provide an input-output relationship from the voltage of the source to the current flowing through the joint.

\begin{figure}[ht!]
    \centering
         \caption{Typical operation in open-loop; $I_W$ is the current through the joint, $E_W$ is the
         source's voltage and $U_{arc}$ is the voltage drop in the
         joint}
        \includegraphics[scale=0.27]{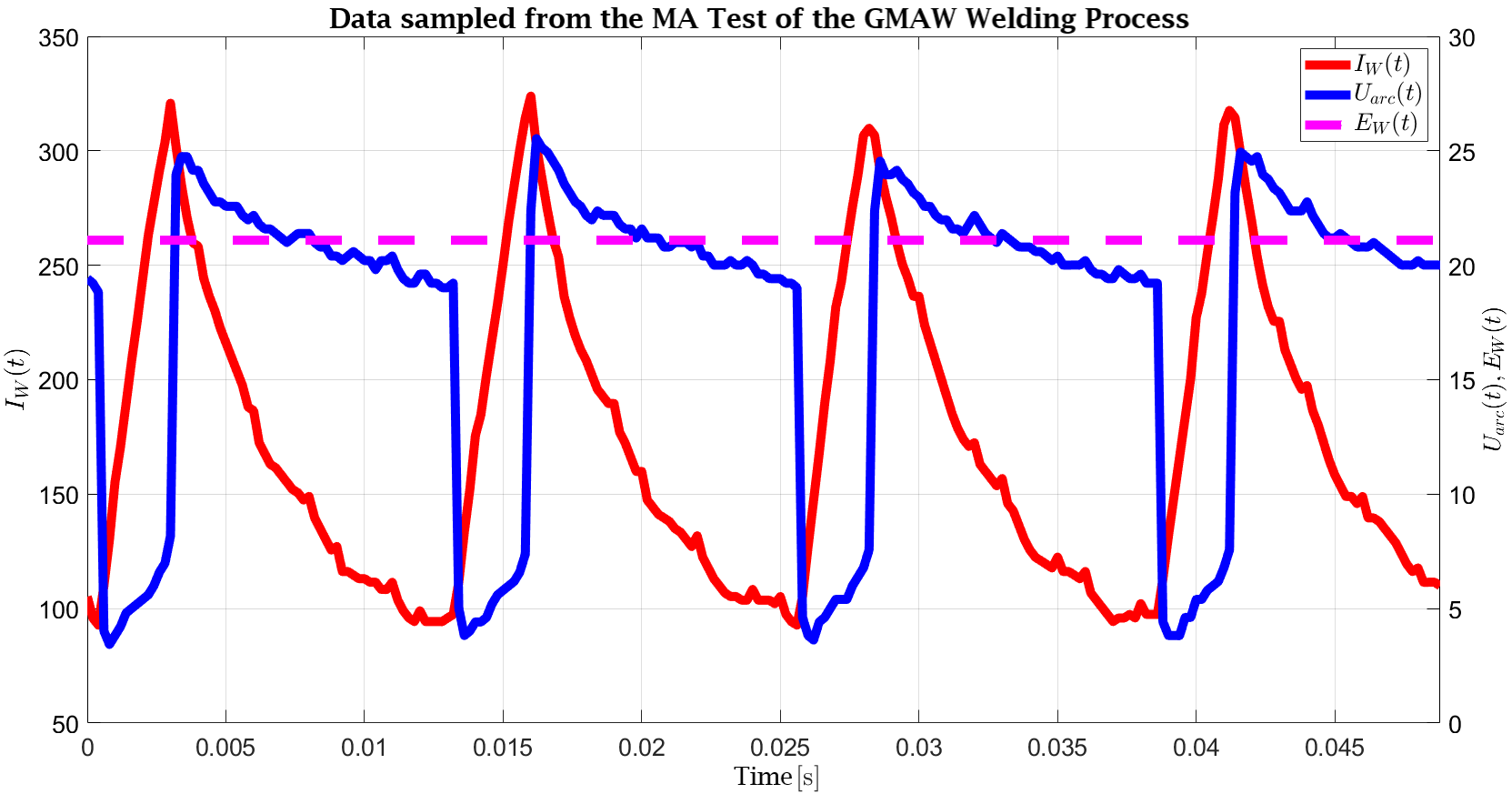}\\
        \label{fig:real_data}
\end{figure}

The starting point to conceive such a model is the observation of the prototypical behavior presented in Figure \ref{fig:metal_transfer}, which is also observed in the real data presented in Figure \ref{fig:real_data}.  In this Figure, $I_W$ is the current through the joint - the process variable, $E_W$ is the source's voltage - the to-be control input, kept constant in this experiment - and $U_{arc}$ is the voltage drop in the joint. It is observed that at each phase the behavior is different, which suggests a switched model. This is also suggested by the physics involved, illustrated in Figures \ref{fig:1} and \ref{fig:metal_transfer}, which show that the system indeed switches from one of a short-circuit at the joint to one in which there is no short-circuit but an electric arc at the joint. On the other hand, it is also observed that in both phases the relationship between input and output resembles the behavior of a passive circuit. Considerations of this nature have previously led to the proposition of an equivalent circuit model, to be presented next.
 
%
%

\subsection{A switched equivalent  circuit model}

\begin{figure}[ht!]
    \centering
         \caption{Switched circuit model proposed in \cite{Gohr:2002}.}
        \includegraphics[scale=0.432]{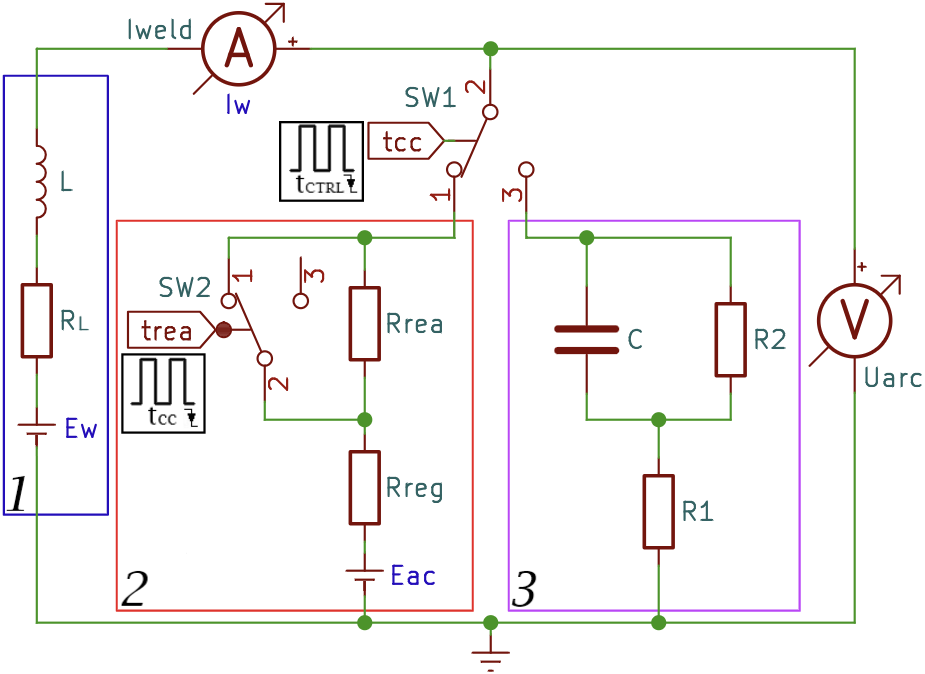}\\
        Fonte: \cite{Gohr:2002}. Adaptado.
        \label{fig:16}
\end{figure}

The following equivalent circuit model, shown in Figure \ref{fig:16}, has been proposed in \cite{Gohr:2002}. The circuit consists of four blocks: block 1, which simulates the welding source by means of a current source; block 2, responsible for simulating the electric arc period; block 3, which represents the short-circuit period; and block 4, intended for measuring the arc voltage. In addition, a function generator defines the frequency of the metal transfer.

%

By applying Kirchhoff's laws to the circuit proposed in Figure \ref{fig:16}, it is possible to describe the relationship between the control voltage $E_{W}$, considered the manipulated input, and the welding current $I_{W}$, which constitutes the controlled output of the system. To do this, the circuits \textit{1} \& \textit{2} and \textit{1} \& \textit{3} are analyzed, from which the transfer functions that represent the dynamic behavior of the process can be derived. Initially considering the circuit \textit{1} \& \textit{3}, which represents the short-circuit phase, one gets the following transfer function:
\begin{align} 
    \frac{I_{sc}(s)}{E_{W}(s)} = \frac{s(CR_{2}) + 1}{s^{2}(CLR_{2}) + s(L + CR_{L}R_{2} + CR_{1}R_{2}) + (R_{1} + R_{2} + R_{L})} .
    \label{math:29}
\end{align}
  
As for the electric arc phase, analysis of the circuit \textit{1} \& \textit{2} yields:

\begin{align} 
    \frac{I_{ea}(s)}{E_{W}(s) - Eac(s)} = \frac{1}{sL + (R_L+R_{rea}+R_{reg})} .
\label{math:32}
\end{align}

%

\subsection{Parameter identification}


The equivalent circuit  model is conceptually meaningful by itself  and allows to analyze qualitatively and simulate the typical GMAW process's behavior. But in order for the model to be useful for control design, adequate parameter values must be obtained that are specific to the process to be controlled. We propose to obtain the appropriate parameter values in a data-driven manner, by means of Prediction Error Identification (PEI) \cite{Ljung:1999}. In so doing, starting with a model structure based on physical principles and estimating its numerical parameters through PEI, we are developing what is usually called a {\em gray box model} \cite{Ljung:1999}. 

In prediction error identification we are given a model set ${\cal M}(\theta)$, parametrized in terms of a real parameter vector $\theta \in \mathbb{R}^d$ for some integer $d$. In our case, we have two model sets, one for each phase. For the short-circuit phase the model set is the set of all transfer functions in the form \eqref{math:29} and the parameter vector is 
$$\theta_{sc} = \left[\begin{array}{c}
			R_1 \\ R_2 \\ C
			\end{array}\right] 
$$
as $L$ and $R_L$ are known a priori. For the electric arc phase the model set is the set of all relationships in the form \eqref{math:32}, noticing that the arc voltage $E_{ac}$ is a constant to be estimated, so the parameter vector is $$
\theta_{ea} = \left[\begin{array}{c}
			(R_{rea} + R_{reg}) \\ E_{ac} 
			\end{array}\right] .
$$

 
With the model set and a batch of input-output data, we form a predictor of the model's output, which depends on the parameter $\theta$, then find the value of $\theta$ for which the predictor's output is most similar to the measurements made. This procedure is formalized below. 

\begin{problem}\label{prob_PEM}
Given a model set ${\cal M}(\theta)$ 
and measurements $E_W(t)$, $I_W(t)$, $t=1, \ldots, N$,
find a  solution $\hat{\theta}$ of the optimization problem:
\begin{eqnarray*}\label{optimization_PEM}
&& \hat{\theta} = arg~min_{\theta} J_N(\theta) \\
&& J_N(\theta) =\frac{1}{N}  \sum_{t=1}^{N} [\hat{I}_W(t,\theta)  - I_W(t)]^2 
\end{eqnarray*}
where $\hat{I}_W(t,\theta)$ is the output of the current predictor.  \end{problem}

It remains to define what the predictor is. The optimal predictor is given by the expected value of the output given previous measurements, that is,
\mbox{$\hat{I}_W^*(t,\theta) = E [I_W(t) ~ | ~\theta , I_W(t-1), I_W(t-2), E_W(t-1), E_W(t-2), \ldots ]$} \cite{Ljung:1999}.
If the measurement error is white, then the optimal predictor is obtained simply by simulation of the model and this is the predictor we have adopted in this paper. 

An open-loop data acquisition test was conducted 
on the SUMIG inverter welding power source illustrated in Figure \ref{fig:WeldingPS}. The experiment was performed with the welding power source operating in the short-circuit metal transfer mode, using C15 shielding gas (85\% Argon and 15\% Carbon Dioxide) at a flow rate of 15 L/min, solid wire of 1.2 mm diameter type ER70S-6, and welding parameters of 180 A and 18 V, with a wire feed speed of 5 m/min. The average control voltage signal $E_{W}$ applied was 21.1 V, corresponding to a duty cycle of 10.55\%, considering that the maximum duty cycle is limited to 50\% due to the converter topology. The welding process for data collection was carried out manually.

\begin{figure}[ht!]
    \centering
         \caption{Switched model vs experimental data}
        \includegraphics[scale=0.27]{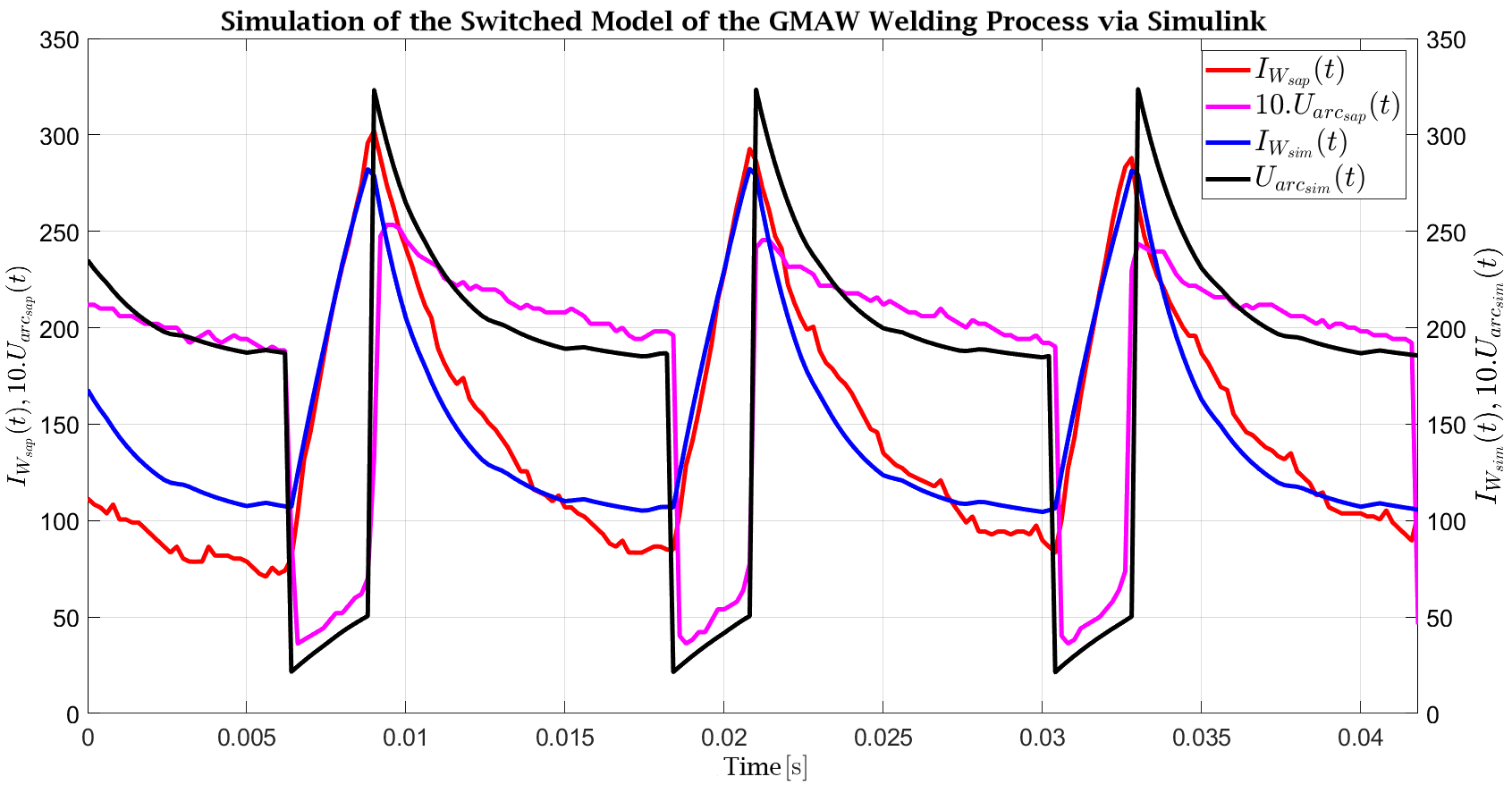}\\
        \label{fig:validation}
\end{figure}

We have solved the  data-driven parameter identification Problem  \ref{prob_PEM} with the data thus collected. Since for each phase we have a different model and a different parameter vector, the PEI has been performed separately for each one of them, as if for two independent systems. The results of PEI are presented in Table \ref{Id_results}, in which the model parameters are all given, and in Figure \ref{fig:validation}, where the real data are compared to the simulation of the model obtained.

\begin{table}[ht!]
    \begin{center}
    \caption{Parameters identified for the switched model}  
    \vspace{2pt}
    \label{Id_results}
    \begin{tabular}{c|cc}
        \hline
         Symbol & Description & Value\\
        \hline
        $L$ & Welding Source Inductance & 180 $\mu$H\\
        $R_{L}$ & Welding Source Resistance & 0,016 $\Omega$\\
        $dT$ & Duty Cycle & 10,55 \%\\
        \hline
        $C$ & Short-circuit Capacitance & 2000 $m$F\\
        $R_{c}$ & Capacitor Resistance & 0,020 $\Omega$\\
        $R_{1}$ & Short-circuit 1 Resistance & 0,010 $\Omega$\\
        $R_{2}$ & Short-circuit 2 Resistance & 0,010 $\Omega$\\
        $R_{rea}$ & Electric Arc Resistance & 0,043 $\Omega$\\
        $R_{reg}$ & Arc Re-ignition Resistance & 0,045 $\Omega$\\
        $E_{ac}$ & Constant Arc Voltage & 11 V\\
        $t_{cc}$ & Short-circuit Period & 2,5 ms\\
        $t_{ae}$ & Electric Arc Period & 9,5 ms\\
        \hline  \hline
        MSE$_{I_{W_{sim}}}$ & \textit{Mean Squared Error} $I_{W_{sim}}$ & 436,79\\
        MSE$_{U_{arc_{sim}}}$ & \textit{Mean Squared Error} $U_{arc_{sim}}$ & 10,59\\
        \hline
    \end{tabular}\\
    \vspace{2pt}
    \vspace{-20pt}
    \end{center}
\end{table}

\section{Control design}
 
The model obtained by the means presented in the previous Section 
will now be used to design a switched PID current controller. But first
the controller specifications must be detailed.

\subsection{Specifications}

It has been established that an appropriate current waveform is \cite{Gohr:2002}:
\begin{itemize}
    \item a particular rate of growth during the short-circuit phase; for our particular process this optimal rate is $60~\frac{A}{ms}$
    \item a particular rate of decay during the electric arc phase; for our particular process this optimal rate is $20~\frac{A}{ms}$. 
\end{itemize}

\begin{figure}[ht!]
    \centering
         \caption{Schematic diagram of the proposed control system.}
        \includegraphics[scale=0.38]{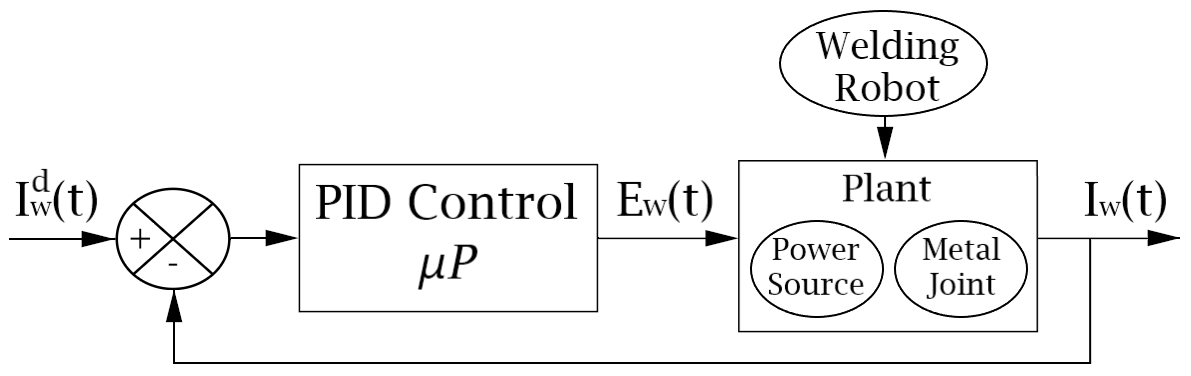}\\

        \label{fig:blocos}
\end{figure}

\begin{figure}[ht!]
    \centering
         \caption{Closed-loop behavior of the proposed control system in Figure \ref{fig:blocos}.}
        \includegraphics[scale=0.25]{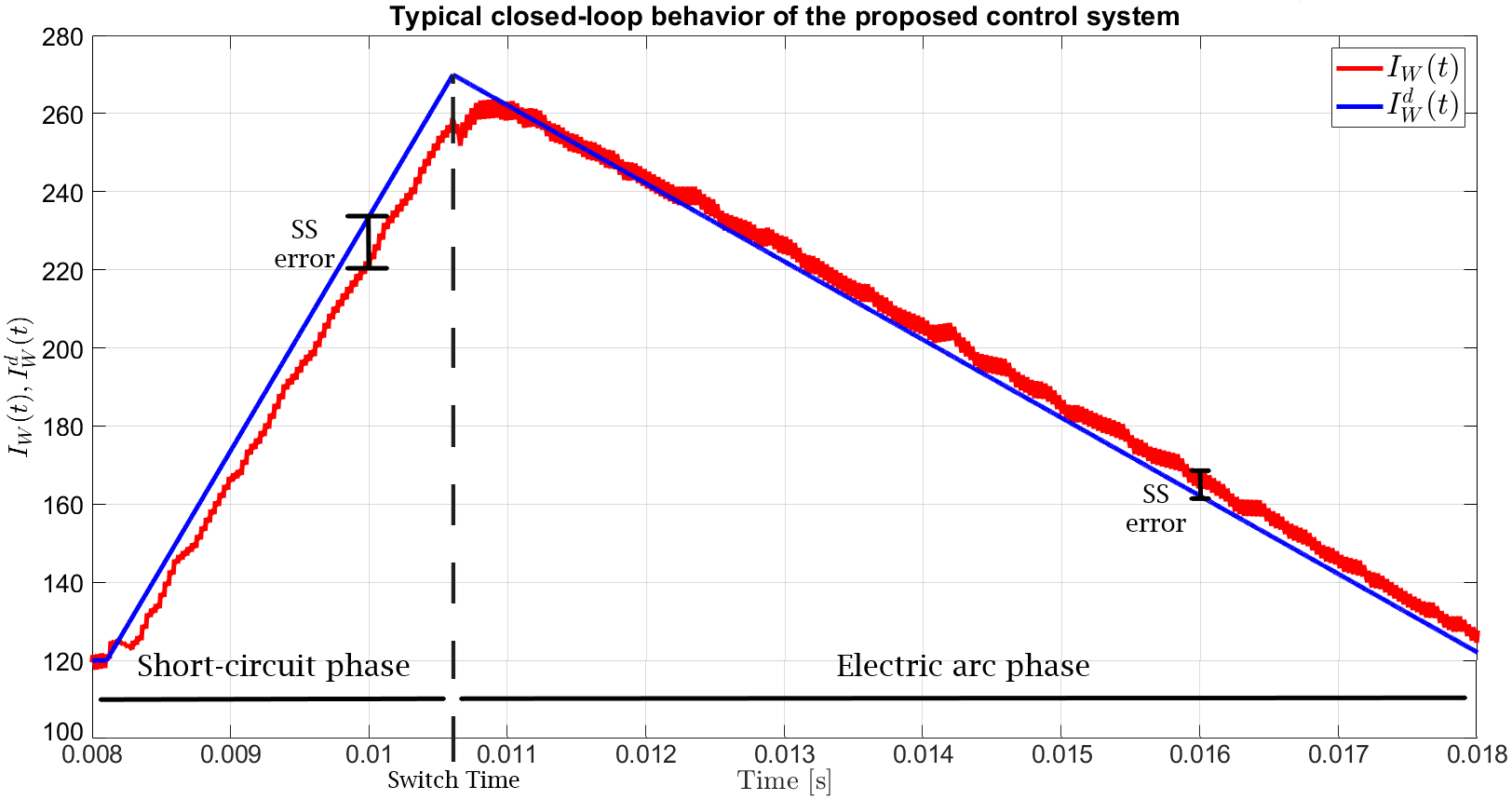}\\

        \label{fig:timeplot}
\end{figure}

These are the specifications for steady-state behavior at each phase of the process. One could thus take the time derivative of the electric current as the process variable and design a controller so that it tracks the desired reference value in each phase. However, this would require to calculate the time derivative of the current, which is almost never advisable from a practical point of view due to amplification of noise involved in this calculation \cite{Ogata}. This solution would be particularly ill-advised in the GMAW process, in which the current signal tends to be very noisy to begin with. To avoid calculation of time derivatives, we propose to use the current itself as the process variable and use as reference signal a ramp with the desired slope. Then the actual current does not need to track the specified ramp reference with zero error, but only with some finite error, as this implies that the slope of the actual current will be the same as that of the reference.
To explain this rationale, a schematic diagram of the control system and a prototypical closed-loop behavior are given in Figures \ref{fig:blocos}
and \ref{fig:timeplot} respectively.

In Figure \ref{fig:timeplot} the behavior of the current in one cycle of metal transfer is shown. In each cycle, at the onset of each phase (short-circuit and electric arc) there will be a transient. The desired  behavior, in which the current $I_W(t)$ behaves like a ramp with the desired slope, will be achieved only after this transient, as can also be seen in Figure \ref{fig:timeplot}. Thus, it is important that the transients last for only a small fraction of each phase  - say $5\%$. Given that the short-circuit phase lasts an average of $2.5~ms$ and the electric arc phase an average of $9.5~ms$, the specifications are that the settling times in closed-loop must be around $125~\mu s$ and $500~\mu s$ respectively.

In summary, the specifications for closed-loop behavior are:
\begin{enumerate}
    \item joint current $I_W$ is the process variable
    \item voltage $E_W$ is the manipulated (control) variable
    \item finite error in steady-state for ramp inputs
    \item settling time smaller than $125~\mu s$ in the short-circuit phase
    \item settling time smaller than $500~\mu s$ in the electric arc phase
\end{enumerate}

Moreover, the desired behavior for the current at each phase of each cycle is given by the reference signal $I_W^d(t) = I_o + \alpha t$, where $I_o$ is the current's value measured at the switching between the two phases and $\alpha$ is the desired rate of change during each phase - $\alpha = 60~\frac{A}{ms}$ in the short-circuit phase and $\alpha = -20~\frac{A}{ms}$ during the electric arc phase, as seen above.

Specification $\#3$ requires an integrator to be included in the controller \cite{Ogata}. As for the specified transient behavior, the required settling times in closed-loop (specs $\#4$ and $\#5$) are not much different than in open-loop, so a controller with second-order dynamics should provide enough freedom to satisfy them. Accordingly, a proportional-integral-derivative controller has been used. 
 
 \subsection{Design of the Proportional-Integral-Derivative controller}
 
A proportional-integral (PID) controller is defined as:
\begin{equation}\label{PID}
E_W(t) = k_p e(t) + k_i \int_0^t e(\tau) d\tau + k_d \frac{de(t)}{dt}
\end{equation}
where $e(t) = I_W^d(t) - I_W(t) $ is the tracking error, $k_p$, $k_i$ and $k_d$ are positive gains to be calculated so that the transient performance is as specified. Equation \eqref{PID} describes an ideal PID controller, whereas a practical implementation must account for the deleterious effects of noise for high-frequency performance. This is accomplished by the addition of a low-pass filter to the derivative action. The resulting transfer function of the PID controller, after addition of this filter and application of the Laplace transform to \eqref{PID}, is given by:
 \begin{equation}\label{PIDs}
 E_W(s) = [k_p + \frac{k_i}{s} + k_d\frac{s}{1+Ts} ]E(s)
 \end{equation}
 where $E(s)$ is the Laplace transform of $e(t)$ and $T$ is the time constant of the low-pass filter.
 
Since the specifications are given explicitly in terms of transient performance, it is natural and easy to approach the design by Root Locus \cite{Ogata}. This is an easy design, since in both operation modes the plant is of low order and the transient specifications are not hard - specified settling times in closed-loop are similar to open-loop. As a result of the Root Locus design the controller's gains are the ones given in Table \ref{gains}.

\begin{table}[ht!]
    \begin{center}
    \caption{Parameters of the switched PID controller }  
    \label{gains}
    \vspace{1mm}
    \begin{tabular}{|c|cc|}
    \hline
           & short-circuit & electric arc  \\ \hline
    $K_p$ & 4,25    & 1,55   \\ 
    $T_i$ & 6,296.$10^{-3}$    & 1,107.$10^{-3}$   \\ 
    $T_d$ & 1,176.$10^{-3}$    & 1,613.$10^{-3}$   \\ \hline
    \end{tabular}
    \end{center}
\end{table}


\subsection{Switching}

At each operating mode one PID controller will be used, and also a different reference signal, so the controller also includes a logic that detects when the plant switches from one operating mode to another - from short-circuit to electric arc and vice-versa. The start of a short-circuit period is detected when the current's gradient becomes significantly positive at the same time that the arc voltage becomes lower than a previously chosen threshold of $14.4 ~V$. When this happens, the  controller parameters are updated to their short-circuit values - those in the first column of Table \ref{gains}.  On the other hand, when the current's gradient changes signal to become negative, then it is diagnosed that the system has left the short-circuit to enter the electric arc phase; accordingly, at this moment the  controller parameters are updated to their electric arc values - those in the second column of Table \ref{gains}.

 \section{Experimental results}

We have performed a large number of experimental tests to evaluate the efficacy of the proposed current control strategy. These tests involved both manual welding and welding with a robot. 


\subsection{Manual welding}

\begin{figure}[ht!]
    \centering
         \caption{Graphical results of controller tests 2.}
        \includegraphics[scale=0.27]{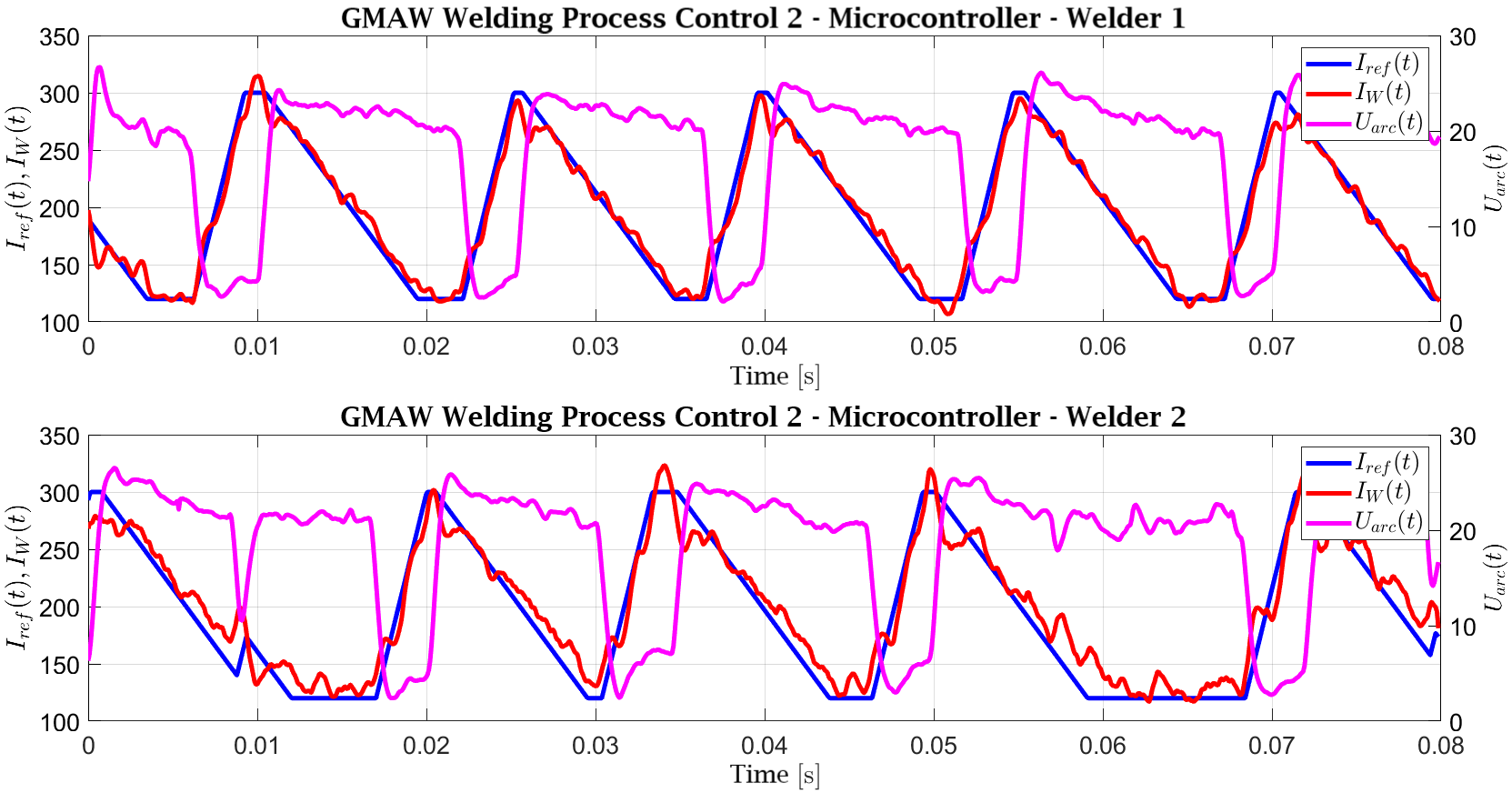}\\

        \label{fig:35}
\end{figure}

Two different welders have welded similar pieces under the same welding parameters, using three different strategies for current control: open-loop, the commercially available closed-loop strategy  and our proposed control strategy. Once the tests have been concluded, we first asked the welders for feedback and they reported that they found the welding experience with our proposed strategy somewhat ``bumpy''. This motivated us to retune the controllers with a 10\% settling time (instead of the 5\% criterion applied initially), which yielded the controller parameters given in Table \ref{gains_bis}. A sample experimental result, showing the typical behavior of the electrical variables in closed-loop with this new controller setting, is presented in Figure \ref{fig:35}. It is seen in this Figure that the current  indeed behaves as specified; closed-loop performance measures are given in Table \ref{gains_bis}. Also note how the arc voltage (the manipulated variable) automatically varies in order to enforce the current's behavior. On the other hand, the welders's feedback was that with this new tuning welding was indeed quite steady. 

Objective measures of performance are presented in Table \ref{tab:7} for both controller settings. It is seen that the control performance was very close with both settings (as expected), but the behavior of the electrical variables (seen in Fig \ref{fig:35}) was indeed smoother with the retuned controller, confirming the welders's feel.

 \begin{table}[ht!]
    \begin{center}
    \caption{Parameters of the switched PID controller }  
    \label{gains_bis}
    \vspace{1mm}
    \begin{tabular}{|c|cc|}
    \hline
           & short-circuit & electric arc  \\ \hline
    $K_p$ & 2,75    & 3,25   \\ 
    $T_i$ & 39,286.$10^{-3}$    & 13.$10^{-3}$   \\ 
    $T_d$ & 454,55.$10^{-6}$    & 769,23.$10^{-6}$   \\ \hline
    \end{tabular}
    \end{center}
\end{table}


\begin{table}[ht!]
    \begin{center}
    \caption{Summary of the performance measures obtained with manual welding}  
    \label{tab:7}
    \begin{tabular}{c|cccc}
        \hline
        Description & C1$_{S1}$ & C1$_{S2}$ & C2$_{S1}$ & C2$_{S2}$\\
        \hline
        $d$I/$d$t$_{S}$ & 55,4 A/ms & 60,6 A/ms & 60,8 A/ms & 59,9 A/ms\\
        $d$I/$d$t$_{D}$ & 22,6 A/ms & 24,7 A/ms & 21,6 A/ms & 21,0 A/ms\\
        I$_{effective}$ & 177,8 A & 176,6 A & 180,1 A & 181,5 A \\
        V$_{effective}$ & 19,3 V & 19,6 V & 20,0 V & 19,3 V\\
        I$_{PEAK_{avg}}$ & 275,4 A & 274,1 A & 274,0 A & 280,5 A \\
        V$_{PEAK_{avg}}$ & 25,4 V & 26,1 V & 25,9 V & 24,8 V\\
        $d$T$_{AE_{avg}}$ & 12,6 ms & 12,6 ms & 10,6 ms & 11,6 ms\\
        D.P. $d$T$_{AE_{avg}}$ & 6,8 ms & 7,4 ms & 5,6 ms & 5,9 ms\\
        $d$T$_{CC_{avg}}$ & 2,5 ms & 2,5 ms & 2,3 ms & 2,5 ms\\
        D.P. $d$T$_{CC_{avg}}$ & 1,5 ms & 1,5 ms & 1,3 ms & 1,3 ms\\
        \hline 
        \end{tabular}\\
        \vspace{0pt}
        \vspace{-10pt}
        \end{center}
\end{table}

\subsection{Robot welding}

The collaborative robot used was the FR10, developed by FAIR Innovation Robot Systems, featuring six degrees of freedom, a general repeatability of 0.05 mm,
and a positional repeatability of 0.03 mm, in accordance with ISO 9283. The process variables of attack angle of 50° and stick-out of $20~mm$ were selected,
and the torch travel speed was set at 25 cm/min for all tests.
%

Nine experiments have been performed: one with open-loop current, one with the commercial closed-loop control and seven with our proposed strategy, six of which with the retuned controller settings in Table \ref{gains_bis} and one with the original controller settings in Table \ref{gains}.
A summary of the performance achieved with robot welding is provided in Table \ref{tab:9}. It is seen that all controllers achieved average values of current rate very close to specified, with 5\% discrepancy in the worst case, whereas in the traditional strategies the discrepancies are very large - around 30\% and above in all cases.


\begin{table}[ht!]
    \begin{center}
    \caption{Compiled from numerical results via SAP V4.40s of robotic welding tests.}  
    \label{tab:9}
    \begin{tabular}{c|ccc}
        \hline
        Description & Worst - best all controllers & Open-loop & \textit{Intellimig500}\\
        \hline
        $d$I/$d$t$_{S}$ & 56,3 - 58,5 A/ms & 
        76,6 A/ms & 43,4 A/ms\\
        $d$I/$d$t$_{D}$ & 23,0 - 20,1  A/ms & 45,2 A/ms & 35,7 A/ms\\
        I$_{effective}$ & 194,7 A & 184,5 A & 176,3 A \\
        V$_{effective}$ & 19,5 V & 19,2 V & 19,3 V\\
        I$_{PEAK_{avg}}$ & 293,6 A & 309,8 A & 268,1 A \\
        V$_{PEAK_{avg}}$ & 24,8 V & 25,2 V & 23,3 V\\
        $d$T$_{AE_{avg}}$ & 9,5 ms & 8,7 ms & 13,8 ms\\
        D.P. $d$T$_{AE_{avg}}$ & 5,2 ms & 2,4 ms & 7,2 ms\\
        $d$T$_{CC_{avg}}$ & 2,4 ms & 2,5 ms & 2,8 ms\\
        D.P. $d$T$_{CC_{avg}}$ & 1,5 ms & 0,6 ms & 1,3 ms\\
        WireFeed Speed & 5,4 m/min & 5,4 m/min & 5,0 m/min\\
        \hline 
        \end{tabular}\\
        \vspace{0pt}
        \vspace{0pt}
        \end{center}
\end{table}

\subsection{Metallographic analysis}
 
\begin{figure}[!ht]
    \centering
         \caption{Manual test specimen controllers 2 - welder 2.}
        \includegraphics[scale=0.115]{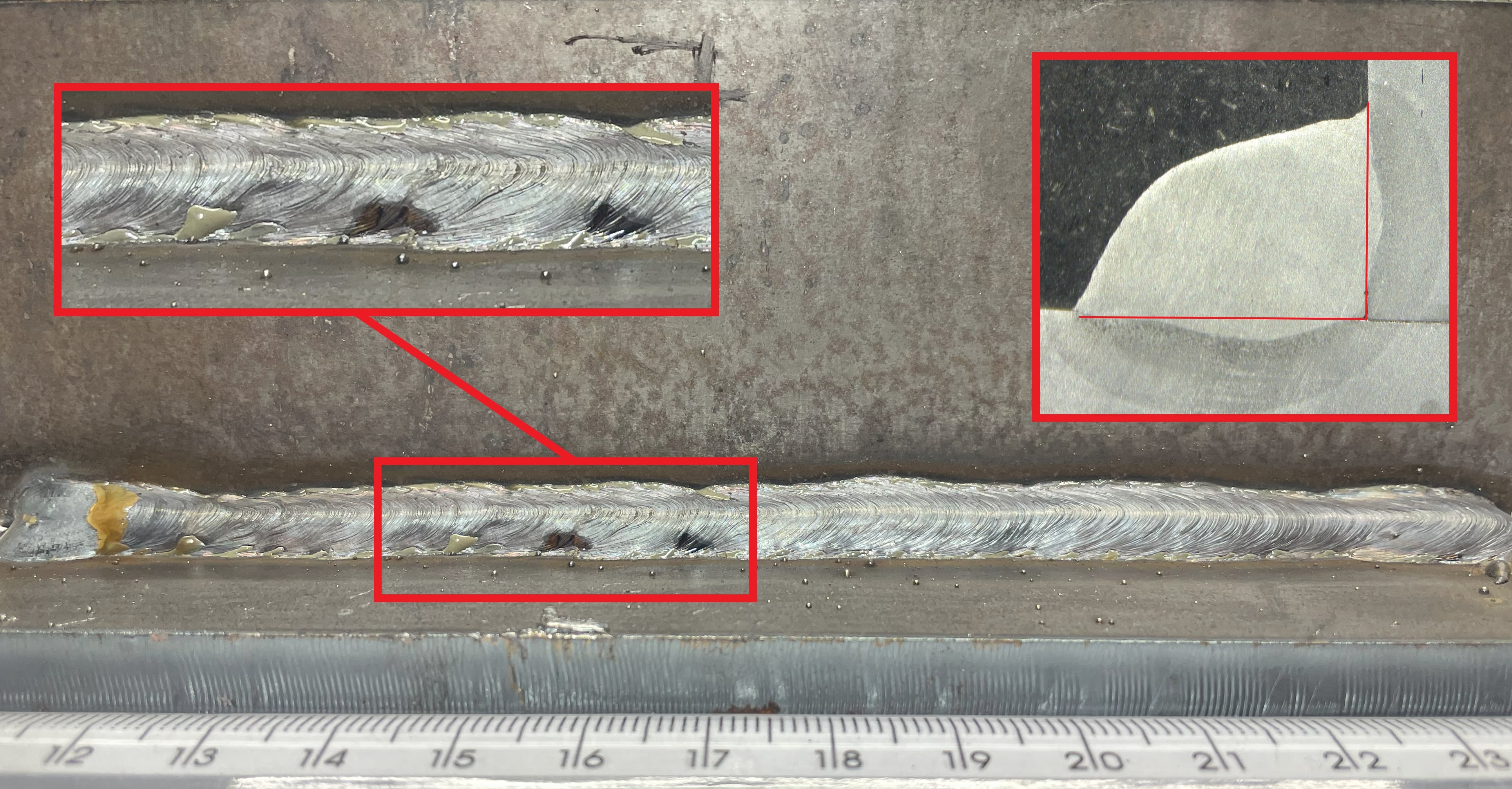}\\
        \vspace{-2pt}
        \label{fig:provaC2}
\end{figure}

\begin{figure}[ht!]
    \centering
         \caption{Robotic test specimen controllers 1.}
       \includegraphics[scale=0.113]{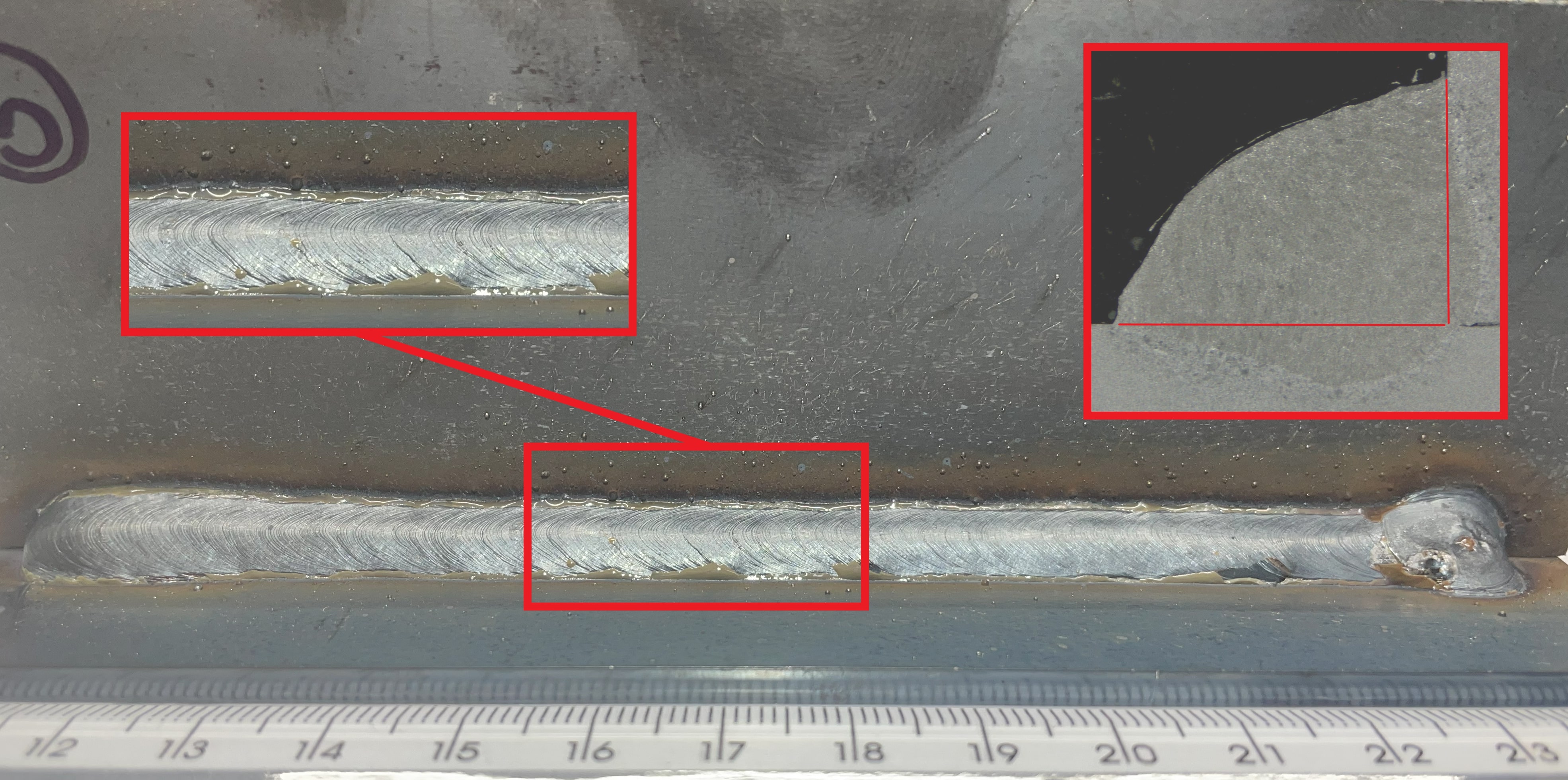}\\
        \vspace{-2pt}
        \label{fig:corpo_robo}
\end{figure}

We have performed metallographic analysis of {\bf 25} pieces welded with our proposed current controller. Sample results are presented in Figure \ref{fig:provaC2} for manual welding and in Figure \ref{fig:corpo_robo} for robot welding. It is seen in these figures that the weld bead is clean, there is little spattering and that there is good penetration in all directions. In fact, it has been observed in the numerous experiments performed that spattering was much reduced with our proposed current control strategy.  
 
 \section{Conclusions}

In this paper we have proposed a particular closed-loop current control to improve the quality of the GMAW process. The proposed  strategy consists of a switched PID controller designed by formal control design methods, based on a switched model for the electrical behavior of the welding joint. We have implemented digitally this switched PID controller and performed numerous tests, both with manual welding and with robot welding. The tests have shown significant improvements in current behavior with respect to traditional current control methods and overall improvement in various performance measures. Metallographic analysis performed in various test specimens have shown high quality of the joints obtained.
Future work will focus on  developing data-driven control design, so that the design is automated and can be made adaptive to varying welding parameters and different conditions resulting from different welders or welder robots; and on the application to enforcing other, more complex current waveforms.


\section*{Declarations}

\subsection*{Availability of data and material}

All data analyzed during this study can be made available under request.

\subsection*{Competing Interests}

The authors declare no competing interests.

\subsection*{Funding}

 This work has been supported by CNPq - Conselho Nacional de Desenvolvimento Cient\'{\i}fico e Tecnol\'ogico - through a personal grant to the first author. 

\subsection*{Authors' contributions}

Both authors together conceived the control system proposed. Alexandre Sanfelici Bazanella checked all the results, wrote the text, produced figures and tables. Mateus Gaspari de Freitas conceived and performed all experiments, collected all data, produced figures and tables, checked the final text and adjusted parts of it. 

\subsection*{Generative AI in scientific writing}

The authors declare
that they did not use generative artificial intelligence (AI) or AI assisted technologies in the writing process.

 \bibliographystyle{IEEEtran}
\bibliography{references_Mateus}

\end{document}